\documentclass[review]{elsarticle}

\usepackage{lineno,hyperref}
\usepackage{xspace}
\PassOptionsToPackage{hyphens}{url}\usepackage{hyperref}

\usepackage{ulem}

\usepackage{amsmath,amssymb,amsfonts}
\usepackage{algorithmic}
\usepackage{graphicx}
\usepackage{textcomp}
\usepackage{xcolor}
\def\BibTeX{{\rm B\kern-.05em{\sc i\kern-.025em b}\kern-.08em
    T\kern-.1667em\lower.7ex\hbox{E}\kern-.125emX}}

\usepackage{graphicx}
\usepackage{eurosym}

\usepackage{float}
\usepackage{placeins}

\usepackage{tabularx}
\usepackage{longtable}
\usepackage{supertabular}

\usepackage{hyperref}

\usepackage{xcolor}

\usepackage{bbding}
\usepackage{pifont}
\usepackage{wasysym}
\usepackage{amssymb}
\usepackage{xurl}

\usepackage{tablefootnote}

\usepackage{multirow}
\usepackage{tikz,lipsum,lmodern}
\usepackage[most]{tcolorbox}
\newtcbtheorem[no counter]{note_Rocio}{R}{
breakable,
lower separated=false,
colback=gray!10, 
colframe=white, fonttitle=\bfseries,
colbacktitle=cyan!20, 
coltitle=black,
enhanced,
boxed title style={colframe=cyan!20},
attach boxed title to top left={xshift=0.5cm,yshift=-2mm},
}{stp_w}

\newtcbtheorem[no counter]{note_ToDo}{TO DO}{
breakable,
lower separated=false,
colback=gray!10, 
colframe=white, fonttitle=\bfseries,
colbacktitle=yellow!20, 
coltitle=black,
enhanced,
boxed title style={colframe=black},
attach boxed title to top left={xshift=0.5cm,yshift=-2mm},
}{stp_w}

\journal{Journal of Parallel and Distributed Computing}

\begin{document}
\newcolumntype{L}[1]{>{\raggedright\arraybackslash}p{#1}}
\newcolumntype{C}[1]{>{\centering\arraybackslash}p{#1}}
\newcolumntype{R}[1]{>{\raggedleft\arraybackslash}p{#1}}

\begin{frontmatter}

\title{Leveraging Teaching on Demand: Approaching HPC to Undergrads}


\author[ucm_address]{Sandra Catal\'an\corref{mycorrespondingauthor}}
\ead{scatalan@ucm.es}
\author[uji_address]{Roc\'io Carratal\'a-S\'aez}
\ead{rcarrata@uji.es}
\author[uji_address]{Sergio Iserte}
\ead{siserte@uji.es}

\address[ucm_address]{Universidad Complutense de Madrid (UCM), Madrid, Spain}
\address[uji_address]{Universitat Jaume I (UJI), Castelló, Spain}
\cortext[mycorrespondingauthor]{Corresponding author.}

\begin{abstract}

High Performance Computing (HPC) is a highly demanded discipline in companies and institutions. However, as students and also afterwards as professors, we observed a lack of HPC related content in the engineering degrees at our university, including Computer Science. Thus, we designed and offered the engineering students a non-mandatory course entitled ``Build you own Raspberry Pi cluster employing Raspberry Pi'' to provide the students with HPC skills. With this course, we covered the basics of supercomputing (hardware, networking, software tools, performance evaluation, cluster management, etc.). This was possible thanks to leveraging the flexibility and versatility of Raspberry Pi devices, and the students' motivation that arose from the hands-on experience. Moreover, the course included a ``Teaching on demand'' component to let the attendees choose a field to explore, based on their own interests. In this paper, we offer all the details to let anyone fully reproduce the course. Besides, we analyze and evaluate the methodology that let us fulfill our objectives: increase the students' HPC skills and knowledge in such a way that they feel capable of utilizing it in their mid-term professional career.
\end{abstract}

\begin{keyword}
Computational Cluster \sep Undergrad Teaching \sep System Administration \sep Parallel and Distributed Computing \sep Raspberry Pi
\end{keyword}

\end{frontmatter}


\section{INTRODUCTION}

The mathematician and computer scientist Alan Turing said ``We can only see a short distance ahead, but we can see plenty there that needs to be done''.  High-Performance Computing (onward, HPC) is nowadays vital for almost every researcher and company, covering from data analysis to computer simulations and artificial intelligence studies. By analyzing {\it the short distance ahead} of us, as graduates from Universitat Jaume I (UJI), we observed that there exists an indisputable gap between the syllabus of the degrees and the HPC needs of the professional environment. We realized that there is a lack of HPC contents in Computer Science studies, and also in other engineering degrees. For this reason we decided to offer a non-mandatory introductory course entitled ``Build your own supercomputer with Raspberry Pi''\footnote{Course website (in Spanish): \url{https://sites.google.com/uji.es/supercomputadorraspberrypi}}. It serves as an introduction to HPC, starting from the basics and using Raspberry Pi devices to build a dummy supercomputer.

In this paper, we present our experience with the two editions of the course we have offered in 2018~\cite{teachingOnDemand} and 2019, respectively. It is common to find courses that focus on a specific HPC area, such as parallel programming, machine learning, cluster design from the perspective of selecting specific hardware or recycling it, etc. 
However, from our understanding, a complementary course of introduction to clusters of computers should show all the possibilities that a supercomputer may provide and let the students decide which aspect appeals more to them. For this purpose, in the course, we opted for covering a wider scope of HPC-related fields, providing the attendees with general basics. 

Three main reasons motivated us to create and design the course: 1) HPC society interest and need is increasing; 2) There exists a lack of HPC related content among the Computer Science syllabus at UJI; and 3) HPC self-learning is complicated. The obvious contradiction that 1) and 2) expose, together with the difficulties associated to autonomously start in this field, evidenced that such a course could be useful for our students. Particularly at UJI, the engineering degrees are studied in the School of Technology and Experimental Sciences (ESTCE), which offers 11 degrees, including Computer Science, and 15 masters. The proposed course was offered to all the students in ESTCE since they are expected to have a technical background. Basic Linux command line knowledge was highly recommended, although it was not a strict prerequisite. 

The main goal of the course is to provide the students with HPC knowledge. This includes not only understanding the basics but also being capable of identifying when supercomputers are needed and how they are built, as well as where HPC is applied. To soften the entrance barrier to this field, the course avoids classical teaching methodologies to let students experiment with actual hardware while enjoying the learning process. It is also part of the course's purpose to show the importance of monitoring and data analysis, as well as showcasing the interaction with a real supercomputer.
Besides, a part of the course is left open to adapt its content to the particular interests of the students. To this end, a ''teaching on-demand'' methodology is leveraged, allowing the attendees to choose in what they prefer to invest this specific time slot. 

 

The main contributions of this work are:

\begin{itemize}
    \item Description, test and evaluation of the ``teaching on demand'' approach.
    \item Implementation of a hands-on experience based on the usage of Raspberry Pi devices to motivate the students.
    \item Increase in the HPC knowledge and interest among engineering students.
    \item Detailed description of the course in such a way that the community can reproduce it.
    \item Analysis of the mid-term impact of the course on the attendees academic/professional development.
\end{itemize}

The rest of the paper is structured as follows: in Section~\ref{sec:background} we give some HPC background, focused on describing the terminology, tools, applications, and hardware that are employed in the course; in Section~\ref{sec:related-work} we describe different works that reflect the effort from the HPC community to provide students with skills in the field; in Section~\ref{sec:methodology} we explain the course methodology, providing details about its motivation, goals, curriculum, design, and the students recruitment and selection criteria, as well as describing its structure and the evaluation of the curriculum; in Section~\ref{sec:discussion} we discuss the results extracted from the evaluation of the course curriculum and the lessons learnt that derive; in Section~\ref{sec:follow_up} we present the performed follow-up  to evaluate the mid-term impact of the course on the attendees; in Section~\ref{sec:conclusions} we provide the conclusions extracted from this work; and in Section~\ref{sec:future-work} we explain the open lines that can be explored as future work.

\section{BACKGROUND}\label{sec:background}
In this section, some background to facilitate the understanding of the paper is provided following.
More detailed descriptions of the employed terms can be found in~\cite{hpcglossary,hpcwiki,hpctennessee,hpcpomona}.

\subsection{Terminology}

A {\it cluster} is understood in the context of the course as a set of independent computers (namely {\it nodes}) which can work collectively. Related to cluster, we present the definition of {\it supercomputer}: a system (usually a cluster) that delivers high computational power. Besides, HPC is presented as the field in charge of solving scientific and/or engineering problems, which are typically complex and costly in terms of time and resources. To achieve high performance and, thus, solve large problems, it is necessary to leverage the coordinated and simultaneous work of different devices. This way, solutions can be computed in a reasonable time, much faster than in personal computers. Precisely, the joint execution performed by several processes concurrently is named {\it parallel processing}, and in case those processes do not share the memory, then it is said to be {\it distributed processing}.

The term {\it performance} refers to the computational power
provided by the cluster. Likewise, {\it scalability} is presented as the property that evaluates 
the ratio between two measurements of the time, each of them associated with different hardware/software configurations.

\subsection{Tools and applications}

The basic setup of the cluster includes the network configuration among the nodes, to enable the remote access through {\it Secure SHell (SSH)}; and installing the {\it Network File System (NFS)}, which provides the cluster with a shared file system.
Moreover, parallel and distributed programming will rely on {\it OpenMP} and {\it MPI} to respectively coordinate the execution of several processes within each node, and through different nodes in the system.

\subsection{Hardware}

As mentioned before, to let the students build their cluster, they are provided with {\it Raspberry Pi} devices. These can be seen as small computers composed of a motherboard that integrates the processor (with ARM architecture), RAM, several ports for input/output signals, and a slot for a memory card. The cheap price, together with its versatility, made us opt for this device.


Of course, production HPC clusters are not based on Raspberry Pi, but more powerful components, although there exist actual low-cost energy-aware clusters based on these devices~\cite{primer,segon,quart}. However, in rough outlines, there exists straightforward parallelism between what students can observe when building the cluster employing these simpler devices and what forms actual HPC supercomputers that allows learning the HPC basics~\cite{tercer,quint}. The similarities and differences between production HPC clusters and Raspberry Pi based clusters are explained to students along the course.

\section{RELATED WORK}\label{sec:related-work}

In the recent past, we can find many efforts from the HPC community to convey the supercomputing philosophy to CS and other engineering students. Within these efforts, we can find platforms to simulate HPC environments and help students to understand the different needs and infrastructures~\cite{easypap}, and extra courses to provide them with essential HPC knowledge that is not officially included in their curricula~\cite{hpcmarathons}. Moreover, there are also remarkable efforts such as~\cite{tcppcv} that aim to establish what a CS student should know about HPC, setting some ``core topics''.

Usually, courses involving supercomputing topics focus on parallel programming languages or paradigms~\cite{Holmes2015, Lopez2018}.
Others~\cite{Datti2015}, explain the process of designing a cluster for educational purposes. Commonly, the resulting cluster is based on Beowulf distributed computing system~\cite{Sterling1995}, useful for recycling deprecated machines in the center. In~\cite{Alvarez2018}, the authors combine the hardware and software experience involving students in the cluster set-up, employing Odroid instead of Raspberry Pi. However, that course follows a strict program addressing the implementation of a fixed suggested problem.
A thorough review of experiences using micro-clusters for educational purposes can be read in~\cite{Adams2017}. In that paper, the authors compile a series of low-cost clusters using different hardware such as Parallela, Odroid, NVidia Jetson, and Raspberry Pi. Furthermore, they enumerate several strategies for engaging students.

Apart from scientific publications, several related project descriptions and experiences are available online, where different authors set up and configure a Raspberry Pi cluster for performing distributed tasks.
An effort in building a Raspberry Pi cluster is reported in~\cite{StevenJ.Vaughan-Nichols2013}, where High-Performance Linpack (HPL) is executed over 32 nodes. In~\cite{EvaHavelkova2018}, the authors implement a bingo where each node corresponds to a cell in a bingo card, and all the nodes collaborate to check if the new number completes a match. Projects like~\cite{teachingPDC} provide an image for a Raspberry Pi 3 system with pre-installed support for clusters and other attractive parallel suites. A more recent project~\cite{PJEvans2020} leverages a new generation of Raspberries, the Raspberry Pi model 4 B~\cite{raspy4b} to build a cluster and execute small distributed Python codes.

From our understanding, a complementary course of introduction to clusters of computers should show all the possibilities that a supercomputer may provide and let the students decide which aspects appeal more to them. For this purpose, we have included a wide variety of topics in the course, providing simplified versions of the ones classically addressed in professional and more complex HPC related manuals~\cite{Eadline2018, Severance2018, ASCCommunity2018}, such as software performance, cluster scalability, networking, CPU frequency, cooling, benchmarking, and scientific applications.

The main difference concerning all the other approaches we have analyzed is the fact that we do not focus on any specific aspect (such as parallelism, hardware description, performance evaluation, users management, etc.), but we try to cover the whole scope of subsystems and functionalities while building and utilizing an HPC system, to provide a general HPC overlook.

\section{METHODOLOGY}\label{sec:methodology}

In this section, we describe the proposed course. First, we present the course goals, a high-level curriculum description, the course design, and the recruitment and selection criteria; then, details regarding the contents of the course are provided and discussed; lastly, we include an evaluation of the syllabus.

\subsection{Overview}

Broadly speaking, the course is a 10-hour workshop highly focused on a hands-on approach to bring HPC to students. The main idea is that students, organized in groups, build and configure their cluster. That prototype is used to run HPC applications, to face first-hand issues of supercomputer management, and to experience how to test and characterize a system. 

\subsubsection{Course motivation}
    
The main reason that inspired us to create this course is to bring HPC to undergraduate students. Although it is a cross-cutting subject that is present in a wide variety of fields (for instance physics, economics, or medicine), HPC is far away to be well known by society and, especially, CS and engineering students. In this scenario, three reasons drove us to the creation of the course:

\textbf{HPC society interest and need are increasing.}

The amount of CS professionals that require HPC skills is increasing due to the importance of this area in many social and professional fields. In fact, European and American governments and institutions have increased the funding of HPC related projects~\cite{FinancingHPC_Europe, Obama_HPCInitiative,BenchmarkingHPC_CouncilComptetitiveness}. 

On the other hand, the effort performed by institutions to bring HPC and supercomputers to society, in general, are more common every day. For instance, documents~\cite{bsc_public} and~\cite{tacc_public} are good examples of research centers opening their facilities to the public, to increase the awareness of HPC, supercomputers, and science.

\textbf{There exists a lack of HPC-related content among the Computer Science syllabus.}

There are several previous works in which this fact is also highlighted and addressed~\cite{teachingHPC_UPC,teachingHPC_Texas}. The CS degree of UJI is four years long and includes, basic training (BT), compulsory (C), and optional (OP) subjects, as well as a final degree project (FDP), but the presence of HPC is marginal in the curriculum. The set of subjects included in UJI's Bachelor in CS is described in Table~\ref{Table:CSBSubjects}. 
It can be observed that only 11 out of 62 subjects include HPC-related content. Moreover, three of them are optional, which means that part of the students will conclude their studies having dealt with HPC in less than 13\% of the subjects. Besides, six of the subjects studied during the fourth year correspond to the specialty chosen by the student, and the only two that include HPC content belong to a single specialty.

\begin{table}[hptb]
\centering
\begin{tabular}{|c|c|c|c|}
\hline
\textbf{Year} & \textbf{Type} & \textbf{\#Subjects} & \textbf{\#HPC Subjects} \\ \hline \hline
\multirow{2}{*}{1}            & C             & 9                  & 1                  \\ \cline{2-4}
              & BT            & 1                  & 1                  \\ \hline
\multirow{2}{*}{2}            & C             & 3                  & 0                  \\ \cline{2-4}
              & BT            & 7                  & 3                   \\ \hline
\multirow{2}{*}{3}            & C             & 9                  & 1                  \\ \cline{2-4}
              & OP            & 7                  & 3                  \\ \hline
\multirow{3}{*}{4}             & C             & 24                 & 2                 \\ \cline{2-4}
              & OP            & 1                  & 0                 \\ \cline{2-4}
              & FDP\tablefootnote{It could happen that an HPC related FDP is developed. However, this is not the most common scenario.}           & 1                  & 0                  \\ \hline
\end{tabular}
\vspace{3mm}
\caption{Subjects of the UJI's Bachelor in Computer Science syllabus summarized by year, specifying their type, the total number of subjects which are studied that year, and the number of subjects that include HPC related content (according to those offered in years 2018/2019 and 2019/2020).}
\label{Table:CSBSubjects}
\end{table}

\textbf{HPC self-learning is complicated.}

Subjects such as ``Computer Structure'' and ``Computer Architecture'', in which the basis of HPC are established, typically present the lowest marks among the students, and their failure rates are high. Besides, even though there exist platforms to supplement conventional lectures \cite{Chaudhury2018}, or detailed and well-described manuals~\cite{hennessypatterson}, they often turn out to be complex to understand from the students perspective. Furthermore, a holistic view of HPC requires knowledge from different fields that are unlikely taught along with the CS/OE syllabus, such as configuring applications setup, or job management. All in all,  we consider that self-learning HPC can be a challenging endeavor.

    
\subsubsection{Course goals}\label{sec:course_goals} 
    
When designing the course, we chose its content to guide the students to reach the following main learning objectives:
\begin{itemize}
\item \textit{Objective 1: Gain general HPC knowledge}. We consider that it is crucial to know not only what HPC and supercomputers are, but also what are cutting-edge trends in this field.

\item \textit{Objective 2: HPC not only in supercomputers.} Change their mind about thinking of HPC restricted to huge supercomputers that belong to powerful companies. HPC can be carried out on small infrastructures, such as a personal computer with dedicated hardware and specific software.

\item \textit{Objective 3: A better grasp of supercomputers.} Be able to understand the needs (both in terms of hardware and software) of a supercomputer employed in HPC.

\item \textit{Objective 4: When to use HPC.} Recognize current applications where HPC is necessary.

\item \textit{Objective 5: Enjoy the learning process}. To avoid traditional lessons pressure feelings, flexible methodologies are employed.

\item {\it Objective 6: Relevance of monitoring}. Emphasize the importance of monitoring metrics. For instance, temperature and its impact on the cooling requirements.

\item {\it Objective 7: Analyze performance.} Learn how to conduct and understand performance analysis and scalability.

\item {\it Objective 8: Real-world HPC experience}. Comprehend how an actual HPC cluster/system is built and how users interact with it.

\end{itemize}

    \subsubsection{High-level curriculum overview and its design}
    
\begin{figure}[ht]
    \centering
    \includegraphics[width=\textwidth]{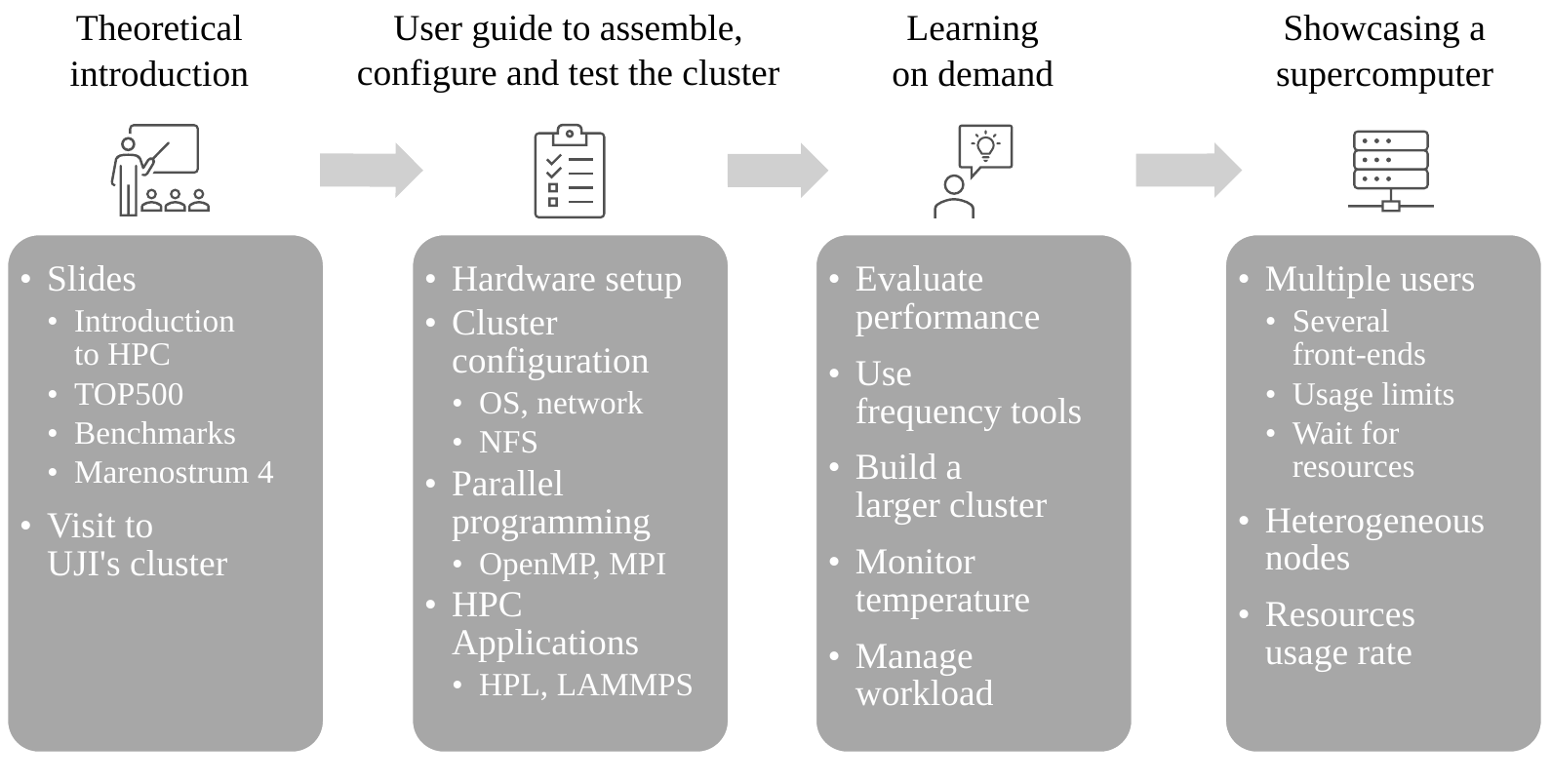}
    \caption{Course curriculum overview.}
    \label{fig:course_overview}
\end{figure}

    The aim of the course was not only to bring HPC to students but also to do it in a practical way. The idea was trying to deviate from traditional lectures to implement a hands-on approach while fulfilling the personal interests of the students. Combining all these goals led us to design the course divided into four stages:
    \begin{itemize}
        \item Theoretical introduction. 
        \item Cluster assembling, configuration, and test.
        \item Learning on demand.
        \item Showcasing a supercomputer.
    \end{itemize}
Figure~\ref{fig:course_overview} illustrates the four stages. Firstly, we provide a theoretical introduction supported by a set of slides. After that, the students are explained how to assemble and configure their cluster. Once the basic configuration and tests are performed, which closes the first part of the course, each group of students can choose on which topic (or topics) wants to focus. Each group works on its chosen topic and, at the end of this stage, results are shared and discussed among the students from all groups. The last part of the course consists of the showcase of an actual supercomputer to give the students a realistic experience as supercomputer users.

    \subsubsection{Design of the course}
    
    A crucial aspect when creating a course is the number of available resources in terms of material, budget, and people. In this course, we propose using a low-cost cluster to guarantee a hands-on experience to the attendees, while keeping a reasonable budget. The required hardware to assemble each of the clusters is described in Table~\ref{Table:Budget}. Hardware peripheral devices such as screens, keyboards, and mice (and the associated wires) were already available in the classroom. This is why they are not included in the cluster budget, although they were employed. 

\begin{table}[H]
\centering
\resizebox{8.5cm}{!} {
\begin{tabular}{|l|c|c|}
\hline
\textbf{Component} 		& \textbf{Quantity} 	& \textbf{Unitary price}  \\ \hline \hline
Raspberry Pi 3 Model B+ 	& 4 					& 29.47\euro \\ \hline
USB Hub (4 ports) 		& 1 					& 11.99\euro \\ \hline
USB 2.0 wires 			& 4 					& 0.94\euro \\ \hline
Micro SD Class 10 (16GB) 	& 4 					& 7.77\euro \\ \hline
Switch Ethernet (5 ports)	& 1 					& 16.50\euro \\ \hline
Ethernet wires 			& 4 					& 1.14\euro \\ \hline \hline 
\multicolumn{2}{|c|}{\textbf{Total price}} 	& 185.77\euro \\ \hline 
\end{tabular}
}
\vspace{3mm}
\caption{Description of the components required for each cluster detailing their prices and quantities (based on year 2018 prices).}
\label{Table:Budget}
\end{table}

In our case, the presented hardware was acquired by the HPC\&A\footnote{http://www.hpca.uji.es/} research group from UJI. Given the available budget and the total cost of each cluster, we could offer 20 vacancies in the course in its first edition, and 24 in the second. The attendees were grouped in a maximum of four-people teams to guarantee the participation of all the members as much as possible.

On the other hand, the amount of human resources is essential in this course, especially to ensure an enriching experience in the on-demand learning part. According to our experience, we consider that one person can be in charge of doing the theoretical introduction. However, the hands-on part requires at least two people in order to help all the groups, given that the duration of the course is restricted to 10 hours. Nevertheless, although the on-demand learning part can be managed by two instructors, we consider that three people are the most recommendable number, since progression pace may vary among the different groups. 
One lecturer takes care of the groups that need more time to complete the user's guide configuration. (Note that is important to take into account that some students are not from CS and also that there are ``early-stage'' CS students that do not count with the background related to building and configuring the cluster.) Meanwhile, another lecturer coordinates all the groups that are interested in creating a larger cluster.
The third instructor guides the groups interested in the performance-energy consumption and helps them using specific tools.
This distribution of the workload among the lecturers is based on our personal experience of the course and can be of course adjusted to the different needs, agendas, budgets, and course vacancies in other contexts.

Regarding the agenda, it must be noted that, for the first edition, we proposed the course as a 2-day workshop (of five hours duration each day), while for the second edition we decided to offer a single day experience that occupied a Saturday morning and afternoon\footnote{Remark: at UJI, there are no lectures on Saturdays.}. The reason to change the format of the course was the students' availability. The second edition was conducted closer to the exams period and, from our experience, we considered it easier for students to find a time slot for one (longer) session than for two (shorter) sessions divided into two different days.


    \subsubsection{Recruitment and selection criteria} 
    Participants enrolled voluntarily in the course ``Build your own supercomputer with Raspberry Pi''. The moderate number of vacancies in the course was constrained by the budget and available facilities. However, the interest in the course was reasonable. In the first edition,
26 students applied for the course in total, 20 were selected as participants, and 18 attended the course. In the second edition, the number of applicants raised to 48, 24 were selected to participate, and 20 attended the course.

HPC inherently belongs to the CS field, however, its interdisciplinary nature and the general approach considered in the course made us offer it to all the engineering students.
To select attendees, we followed a simple idea: as we believe the lack of HPC and supercomputing knowledge should be corrected as soon as possible, we prioritized the students who were in an early stage of their studies (first and the second year). Since the course begins on the basics, only basic Unix and Shell knowledge prerequisites were recommended. Once early-stage students were selected, if there were still vacancies, they were assigned to third and fourth-year students.

To analyze the course's findings and outcomes, we divide the attendees into CS and \textit{other engineering} (OE) students. The distribution of students, according to their studies, is shown in Table~\ref{tab:participants}. Note that, to maintain the proportion in terms of applicants' interest and background, we selected a similar amount of participants from CS and OE, always following first-come, first-served criteria regarding their application.

\begin{table}[htbp]
\centering
\begin{tabular}{|c||c|c|c|c|c|}
\hline
&\multicolumn{2}{c|}{First edition} &\multicolumn{2}{c|}{Second edition} \\ \cline{2-5}
             & CS   & OE    & CS   & OE\\
\hline
Applicants   & 15 (58\%) & 11 (42\%)  & 24 (50\%) & 24 (50\%)\\
Selected     & 12 (60\%) & 8 (40\%)  & 13 (54\%) & 11 (46\%)\\
Attendees   & 12 (66\%) & 6 (33\%)  & 10 (50\%) & 10 (50\%)\\
\hline
\end{tabular}
\vspace{3mm}
\caption{Participants classification per field (Computer Science - CS, Other Engineering - OE), degree of participation (Applicants, Selected applicants, Attendees), and course edition (First, Second).}
\label{tab:participants}
\end{table}

A well-known and widely studied concern in engineering and CS students is gender unbalance. UJI is not an exception and, unfortunately, there is a small percentage of female students. While in the first edition of the course, we had a  female attendee (out of the two applicant women), in the second edition, the number of interested women in the course raised to five. Three of them were selected to participate, but none attended it.


\subsection{Course Curriculum} 
As it has already been stated, the main goal of the presented course is to introduce HPC to CS and OE students at UJI to complement the syllabus of their degrees.

For this purpose, a specialized curriculum has been tailored to provide the expertise needed to fill the gap of knowledge between HPC and other related disciplines. In this sub-section, the curriculum of the course is detailed, and linked to the course objectives introduced in Section~\ref{sec:course_goals}.

\subsubsection{Step 1: What's HPC?}
The course starts with a brief description of the concept {\it HPC}.
Supporting the explanation with schemes and videos, in this part, the following questions are addressed:

\begin{itemize}
\item \textit{Why is supercomputing necessary?}
A series of well-known examples such as weather prediction computations, social network analysis to personalize adverts, voice recognition in smartphones, or traffic information in Google Maps application, are raised to make the students think about the need for extremely fast operations processing.
Besides, a wide discussion about treating vast amounts of data or generating and manipulating real-time results is promoted.

\item \textit{How is supercomputing different from computation?}
Following, the most relevant terms related to HPC are defined by giving general ideas and fostering students' discussions. It is worth noticing that in-depth explanations are out of the course's scope. 
In order to ease the understanding of the terms and technologies, as long as it is possible, they are compared to more familiar concepts such as the ones associated with the components in a personal computer. 

\item \textit{Where does supercomputing take place?}
The introduction concludes with the presentation of the TOP500 list~\cite{top500}, and the HPL (a Portable Implementation of the High-Performance Linpack Benchmark for Distributed-Memory Computers) benchmark~\cite{linpack,hpl}. In more detail, we leverage this moment to expose the system's features occupying the first position in the list and compare it to the Spanish system \textit{Marenostrum 4}~\cite{web_marenostrum}, which also appears in the TOP500. Finally, intending to give a more down--to--earth approach, the course includes a visit to the university computing data center, so the students can see an actual HPC system.
\end{itemize}

This first part of the course aims to convey the big picture of HPC (\textit{Objective~1: Gain general HPC knowledge}), prove that HPC can be implemented at a small-scale (\textit{Objective~2: HPC not only in supercomputers}), and illustrate which are the target applications of HPC (\textit{Objetive~4: Where to apply HPC}).
Thus, a global vision of HPC is presented to the students in the context of nowadays facilities and systems.

\subsubsection{Step 2: Cluster set-up}
In the second part of the course, the students are grouped in teams of 3-4 people, and the groups are given the hardware components of the cluster, and the student's guide (see~\ref{sec:appendix}). This document describes the necessary steps to end up with a fully functional cluster with support for distributed computation.
In this regard, the cluster set-up is divided into these three parts:

\begin{itemize}
    \item \textit{Cluster assembly.}
Each group is in charge of assembling its own cluster composed of four Raspberry Pi 3 Model B+ devices (Figure~\ref{fig:cluster} reflects a sample of an already assembled cluster). Besides, a switch, the corresponding Ethernet wires, four SD cards, and the power supplies are provided. 

\begin{figure}[htb]
\centering
\includegraphics[width=0.6\columnwidth]{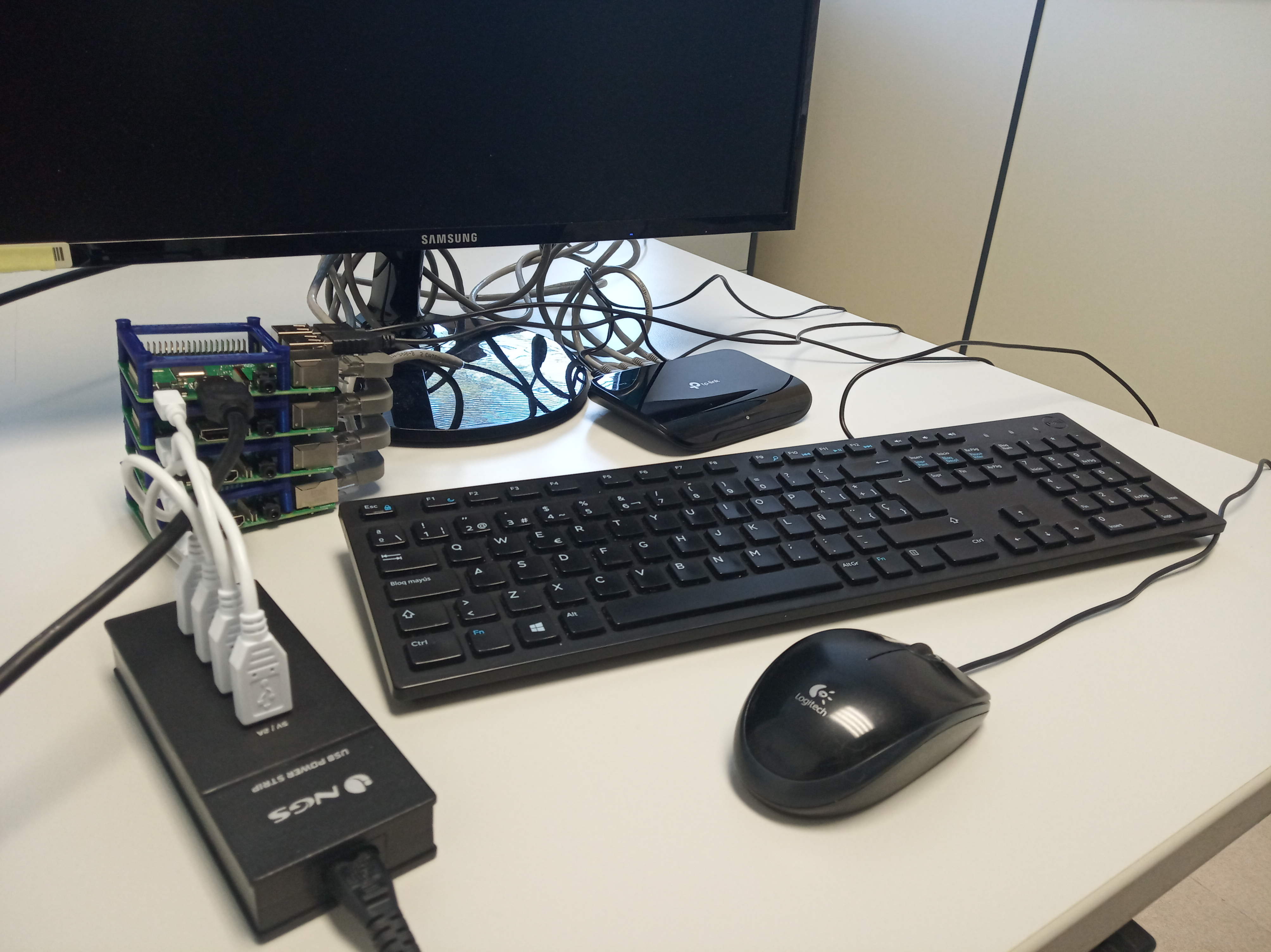}
\caption{A Raspberry Pi cluster assembled during the course. Note that the 3D-printed structure that holds the Raspberry Pi devices is only used for aesthetics purposes.}
\label{fig:cluster}
\end{figure}

Although the guide has already been handed out, the instructors can provide tips and short explanations of the hardware components, especially describing the Raspberry Pi features. Because of the different students' backgrounds or knowledge, with these intermissions, balanced progress among groups is pursued, preventing frustration to the less skilled groups, or boredom to the most advanced ones.

    \item \textit{Cluster configuration.}
The basic software configuration of a cluster comprises the installation of the operating system\footnote{We opted for using Raspbian OS~\cite{raspbian}.}, the network configuration, and the file system set-up. The operating system is pre-installed in bootable SD cards handed out to the attendees to save time and not lose the perspective of the course. Thus, the students can focus on the most relevant steps of the configuration. Once all the components are assembled, the students configure the network and the shared file system. The network configuration is essential to make all the devices work collaboratively. On the other hand, a shared file system 
(Network File System, NFS~\cite{nfs}) is enabled through the cluster to facilitate access to the files within the cluster.

With more detail, students are driven to complete the next steps:

\begin{itemize}
    \item Configure the DHCP service to accept a dynamic IP in the wireless network interface of one single node, which will act as the main node or front-end. Furthermore, the WiFi interface is set as the front-end default gateway. 
    The Ethernet interfaces of all the nodes are configured with a static IP.
    \item Configure SSH by creating public-private pair of keys in the front-end node and then sending the SSH key to the other nodes.
    \item Install and configure an NFS server and create a directory in the front-end shared with all the nodes. It is highly recommended that, at this point, the students check that each node can reach the other nodes, as well as use the NFS properly. For instance, this can be verified by creating a text file located in the shared directory and then editing it from different nodes.
\end{itemize}

Because part of the listed items takes some time to be downloaded, configured, and/or installed, that ``idle time'' is used to explain the corresponding usage and features to the students.

    \item \textit{Performance evaluation.}
To provide a realistic HPC experience, students are introduced to some of the most common software used in a supercomputer.
For this purpose, students will learn how to compile, install and configure:
\begin{itemize}
    \item HPL~\cite{hpl}, a performance benchmark leveraged to rank the supercomputers in the TOP500 list.
    \item LAMMPS~\cite{Plimpton1997}, a classical application for molecular dynamics modeling. 
\end{itemize}
Both applications require the use of the OpenMP programming model~\cite{openmp} and Message Passing Interface (MPI)~\cite{gropp1999using} to exploit all the processing elements in the cluster.
OpenMP is natively integrated into the GCC package and, consequently, no extra installations are required. 
However, MPI needs to be installed. Concretely, we opted for using MPICH~\cite{mpich}, since it is one of the most popular open-source libraries employed in nowadays clusters, along with OpenMPI~\cite{openmpi}.
As we presented both MPICH and OpenMPI to the students, some thought of installing and comparing both implementations, as we will explain in Section~\ref{subsec:MethodPart3}.

The activities proposed for this section are:
\begin{itemize}
\item Configure an execution of HPL to achieve the maximum performance of the cluster. (Note that all nodes should be leveraged.)
\item Perform a scalability analysis of LAMMPS using different configurations of threads and processes.
\item After discussing with the students the importance of also leveraging threads and not only available nodes, we suggest checking the official website, where the way to configure LAMMPS with OpenMP is explained.
\end{itemize}
    
\end{itemize}

 
\textit{Objective~3: A better grasp of supercomputers} and \textit{Objective~4: When to use HPC} are mainly covered during this second part of the course, because the architecture of a supercomputer and how software makes use of its components are explained and related.
Moreover, by means of running benchmarks and helping the students to understand the provided performance results, \textit{Objective~7: Analyze performance} is addressed.


\subsubsection{Step 3: Learning on-demand}
\label{subsec:MethodPart3}
Once the cluster is fully functional, the course is planned to have a free hands-on section, where the students can put into practice the acquired knowledge and/or do research in a specific aspect of their interest, with the help of the instructors.

Although the students can explore any field, they are provided the following on-demand ideas as guidance:

\begin{itemize}

\item \textbf{Performance and frequency tools}.
This activity aims to widen the knowledge about the relationship between performance and energy consumption, the impact of power consumption on HPC facilities, and which tools or strategies can be applied to control these facts.
In this regard, {\it cpufreq} utilities~\cite{Zhou2015ThermalMO} are introduced and tested on their clusters. In this way, students can experience the effect of frequency changes on the performance of executions.

Furthermore, the activity can be complemented with studies of power consumption and the associated economic costs on current supercomputers to illustrate the relevance of the problem. For instance, instructors can provide more information about how performance affects power consumption; dynamic and static power, and their interaction.

\item \textbf{Creating a larger cluster}.
Different teams can decide to merge their clusters to scale their resources up to 8, 12, or more nodes.

The proposed activity requires to reconfigure both the network and the distributed file system in order to ensure the appropriate functioning of the new cluster. With this first step, the knowledge acquired by following the user's guide was reinforced thanks to reproducing that in a new scenario. Once the students check that the new cluster is working correctly, they have the opportunity to try different configurations for LAMMPS and analyze which of them delivers the best performance.

\item \textbf{Real-time measuring of the device temperature}.
Thanks to an incorporated sensor on the board, the temperature can be monitored using the command {\tt vcgencmd measure\_temp}.

This activity requires developing a script that periodically checks the temperature. The data acquired is saved for further analysis.

For instance, Figure~\ref{fig:temperature_students} shows an actual output of this activity provided by one of the groups from the second edition of the course. Concretely, the plot reflects the temperature variations along the seconds required to complete LAMMPS execution setting two different frequencies: 1.40~GHz and 600~MHz. Through this analysis, the students in the group also checked the differences in terms of total execution time when varying the frequency (355 seconds with the higher frequency, in contrast to 661 seconds with the lower one).

\begin{figure}
     \centering
          \includegraphics[width=0.8\textwidth, trim={0 0.4cm 0.4cm 0}, clip]{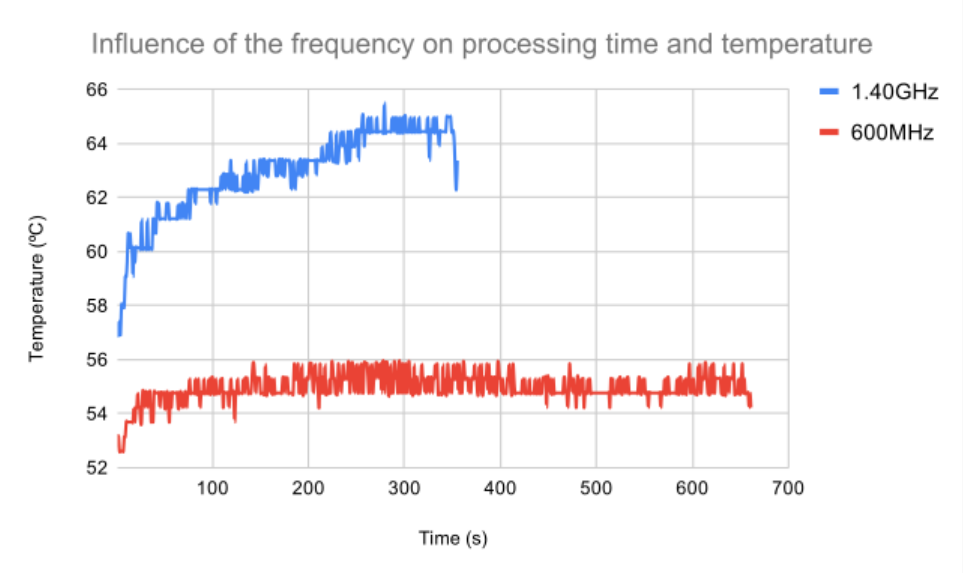}
     \caption{Temperature during a LAMMPS execution with different frequencies.}
     \label{fig:temperature_students}
 \end{figure}

 \item \textbf{Workload management in an HPC system}.
Many users compete for the resources available at the supercomputers. When introducing this notion, students are taught that, with workload management software, resources and users' requests are orchestrated.
For this purpose, Slurm Workload Manager~\cite{Yoo2003} is presented. 
In this activity, following Slurm's administrator guide and helped by the instructors, students face the installation and configuration of Slurm. 

\end{itemize}

Before ending this part, each group is encouraged to briefly share with the others their conclusions on the topic explored and the work performed.

\textit{Objective~5: Enjoy the learning process} is attained by promoting this free hands-on approach. The fact that the students propose the work they want to develop during this part of the course lets them feel relaxed (note that the students are not given a final mark at the end of the course) and free to experience whatever they feel curious about. Also, \textit{Objective~6: Relevance of monitoring} and \textit{Objective~7: Analyze performance} are targeted by showcasing the importance of monitoring tools and how to understand them.

However, some activities may take more time than expected and the students could not finish them on time.
Even in this case, students are encouraged to learn the valuable lesson of \textit{easier said than done}. 
For instance, in the second edition of the course, after installing and configuring Slurm, a team found the following error:
\begin{verbatim}
   slurmctld: error: High latency for 1000 calls to 
   gettimeofday(): 1824 microseconds.
\end{verbatim}
The solution seemed straightforward for the course purposes: recompile the source code increasing the latency tolerance. 
This is translated into an amount of time that is not available in the proposed course.
A {\it failed} activity can be considered a very interesting example to reinforce the experience of the existent difficulties that HPC system administrators, or users, need to solve, and reinforces {\it Objective 8: Real-world HPC experience}.



\subsubsection{Step 4: How to use a supercomputer?}\label{subsec:MethodPart4}
The course wraps up showcasing a real user experience when logging into a top-class supercomputer. This activity demonstrates that large-scale systems have more complexity than the cluster configured in the course. However, the students realize that the cluster philosophy is the same they have learned in the course. It is crucial that, before ending the course, students apprehend the big-picture of a production supercomputer, since they are likely to be the next generation of users, developers, and/or administrators. In this regard, the lecturer in charge logs in a production supercomputer via SSH and explains that:
\begin{itemize}
 \item Unlike the cluster set up in this course, which only has one user and a front-end that, in turn, is a compute node, it is usual to have several dedicated front-ends to provide fault tolerance while supporting hundreds of users interacting with the system simultaneously. For this purpose, the lecturer accesses some of the front-ends and counts the logged users at that moment with the command \texttt{w | wc -l}.
\item All these users share the supercomputer, but they cannot use the hardware in their own free will. We introduce here that there is a management software responsible for temporarily assigning resources to users. Furthermore, it is important to note that users have quotas of usage, which limit the computation time or disk space that a user can utilize.
 \item Resources are not usually idle waiting for us. In production supercomputers, there is a waiting time before assigning resources to the users' jobs. This time varies depending on how saturated the system is, in other words, the number of pending jobs in the queue which are not being executed. In general terms, jobs requesting more resources for more time experience higher pending times.
\item The nodes of the supercomputer do not need to be identical. Supercomputers can be heterogeneous to meet more necessities. In this regard, there exist different partitions of hardware for different types of users.
 Another important concept to bear in mind is that there usually exists a partition with shared machines that are likely to be assigned earlier to the jobs.
\item The employment of the hardware tends to 100\%. For this purpose, we show utilization rate statistics through time. These statistics reflect not only the constant high utilization but also total and partial outages that may indicate maintenance or unexpected failures.\newline
 \end{itemize}

 \indent
 All in all, with this holistic instruction, we are confident that students understand that what they have done is the cornerstone of an HPC system. Hopefully, after leaving the classroom, they will be eager for a second part of the course to learn more about jobs, queues, shared resources, or distributed computation.

 A more realistic experience has been given covering {\it Objective 8: Real-world HPC experience} in this part. This is one of the significant improvements of the second edition in contrast to the first one, made thanks to the change of format between editions, allowing us to include more material that enriches the experience of the attendees by setting them in front of a real cluster in production.


\subsection{Evaluation of the curriculum} 
The new teaching approach presented in this paper to let students become closer to HPC implied risks and challenges. We wanted to be able to judge the impact of this ``custom-tailored'' teaching methodology on students. For this purpose, we collected qualitative information at the beginning and the end of the course through two anonymous surveys. In this section, the questions of the surveys are presented, and those related to the impact of the course on the attendees are analyzed. 

\subsubsection{Initial and final surveys}
The initial and final surveys targeted the potential changes in the attendees' knowledge about HPC. Note that, in the second edition of the course, new questions were included to better justify some of the observations after the first edition. Those questions are marked with the letter ``N''.

We have categorized all students' survey answers to analyze the impact of the course in the short-term (the categories are presented in Figures~\ref{fig:Q2}-\ref{fig:Q5}). The questions were designed to find out how students felt and what they knew about HPC before the course, and what was the progress in terms of HPC knowledge after it.  

Regarding the surveys, the initial one (IS) consisted of the following questions:

\begin{itemize}
\item ISQ1: Why have you signed up for this course?
\item ISQ2: Do you think that HPC has an influence on your day to day? 
\item ISQ3: How would you define HPC?
\item ISQ4: What do you think about supercomputers?
\item ISQ5 (N): In which field do you see yourself developing your professional career when you finish your studies? (Multi-answers were accepted in this question.)
\item ISQ6 (N): Do you believe that HPC could be applied in your desired professional career field?
\end{itemize}

The final survey (FS) consisted of the following questions: 

\begin{itemize}
\item FSQ1: Do you feel more/same/less interested in HPC now? 
\item FSQ2: Do you think that HPC influences your day-to-day?
\item FSQ3: How would you define HPC?
\item FSQ4: What do you think about supercomputers?
\item FSQ5 (N): In which field do you see yourself developing your professional career when you finish your studies? 
\item FSQ6 (N): Do you believe that HPC could be applied in your desired professional career field?
\item FSQ7 (N): Evaluate (from ``0 - Very Bad'' to ``10 - Excellent'') the quality of the theoretical explanations previous to the practical activities.
\item FSQ8 (N): Evaluate (from ``0 - Very Bad'' to ``10 - Excellent'') the empathy degree and quality of the help provided during the practical activities resolution.
\end{itemize}

\subsubsection{Results of the surveys}

\begin{figure}[H]
    \centering
    \includegraphics[width=\textwidth]{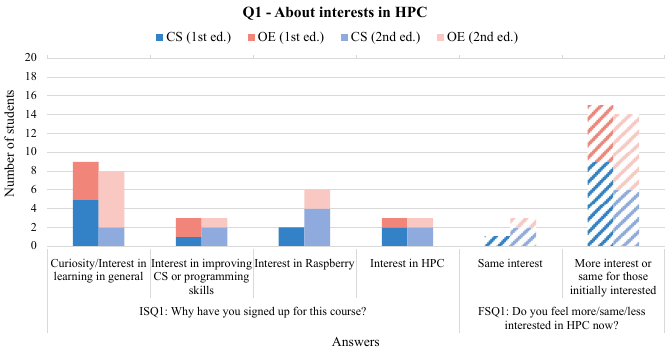}
    \caption{Answers of ISQ1 and FSQ1.}
    \label{fig:Q1}
\end{figure}

\begin{figure}[htbp]
    \centering
    \includegraphics[width=\textwidth]{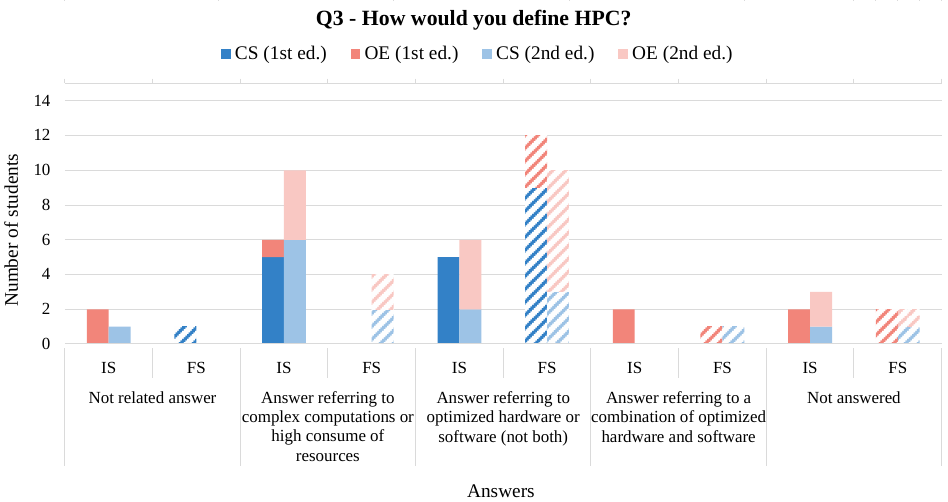}
    \caption{Answers of ISQ3 and FSQ3.}
    \label{fig:Q3}
\end{figure}

\begin{figure}[htbp]
    \centering
    \includegraphics[width=\textwidth]{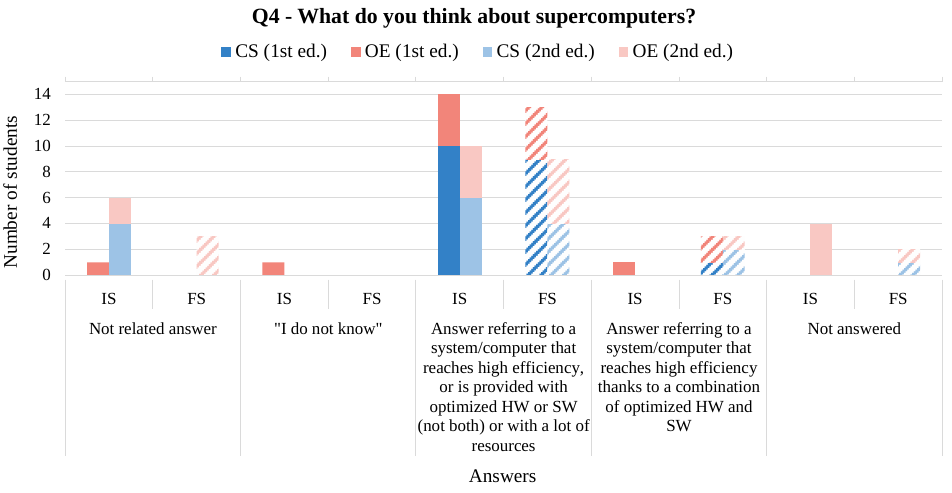}
    \caption{Answers of ISQ4 and FSQ4.}
    \label{fig:Q4}
\end{figure}

\begin{figure}[htbp]
    \centering
    \includegraphics[width=0.85\linewidth]{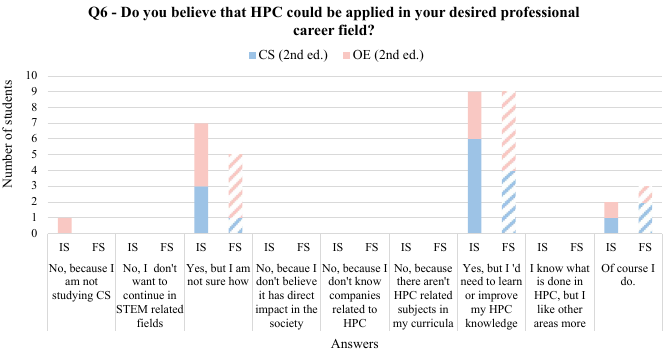}
\caption{Answers of ISQ6 and FSQ6. Note that only 2\textsuperscript{nd} edition answers are provided, as this question was newly included in it.}
    \label{fig:Q6}
\end{figure}

The already categorized answers collected from the students through the surveys are reflected in Figures~\ref{fig:Q1}-\ref{fig:Q7-Q8}. Each plot shows the Initial (IS) and Final Survey (FS) answers to a given question (specified in the title). Data is shown in full colored bars (for IS) and line-filled bars (for FS), including Computer Science (CS) students in blue and Other Engineering (OE) students in red. Differentiation between the first and the second edition answers is made using, respectively, darker and lighter color-schemes. 

Next, we analyze those survey questions that refer to the curriculum evaluation 
(Questions 1 to 4, and Question 6).  
In general, results show that students increase their awareness about HPC in daily life. Moreover, they show good comprehension of usual terms in the HPC field. The main conclusions extracted from the curriculum evaluation are now highlighted. The remaining questions (regarding the course evaluation) are addressed in Section~\ref{sec:discussion}.

\begin{figure}[ht]
    \centering
    \includegraphics[width=\textwidth]{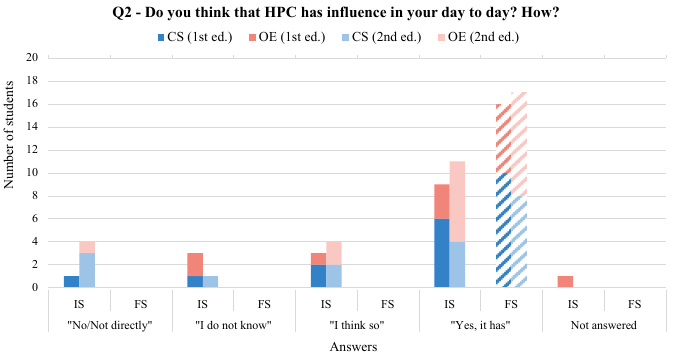}
    \caption{Answers of ISQ2 and FSQ2.}
    \label{fig:Q2}
\end{figure}

\textbf{There exists a lack of HPC knowledge among Engineering students.}

From our teaching experience through the past years, we observed a lack of HPC knowledge among the students. This was an opinion before starting the course, but now we have evidence that exposes that it was true, taking into account what can be observed in Figure~\ref{fig:Q2}. According to this plot, before conducting the course, only around half of the students considered that HPC has an obvious influence daily. It is remarkable that in both editions, around 20\% of the students in total did not know if HPC has that impact (particularly in the first edition, 3 out of 17 attendees), or even negated it (especially in the second edition, 4 out of 20).

\textbf{HPC interest has increased among attendees.}

Regarding the HPC interest and awareness, Figure~\ref{fig:Q1} (FSQ1) shows that, in both editions, the vast majority of students affirm having more interest in HPC than they had before taking the course, regardless of their studies. This is possibly justified by the fact that all of the students in both editions end up the course believing that HPC has an impact on their daily life, which is shown in Figure~\ref{fig:Q2}.

We consider that in case HPC impact in their daily life was more valued, HPC interest in taking the course would have been higher (see Figure~\ref{fig:Q1} (ISQ1)). With that in mind, it seems reasonable that a low number of students applied to the course because of their interest in HPC. In particular, 18\% of the students in the first edition (3 out of 17) and 15\% (3 out of 20) in the second.

\textbf{HPC knowledge has increased among attendees.}

Related to the already mentioned lack of HPC knowledge, we analyze what the students believe HPC and supercomputers are. Figures~\ref{fig:Q3} and~\ref{fig:Q4}, which refer to questions 3 and 4 respectively, illustrate their thoughts. The answers on these plots reflect that defining ``HPC'' or ``supercomputer'' terms implied associating them to ``complex computations'' or ``optimizations'' of the software or the hardware. Those results show that, although their awareness of the daily impact of HPC in our lives is moderate, they have a feeling about what HPC and a supercomputer are even before the course.

Overall HPC knowledge has increased after the course. Figure~\ref{fig:Q3} reflects that in both editions, most of the students abandon their original idea of identifying HPC only with ``solving very complex calculations using lots of resources'' in favor of ``optimizing available software and/or hardware resources''. In the first edition, this percentage increases from 41\% (7 out of 17 students) to 81\% (13 out of 16 attendees), while in the second edition the percentage changes from 30\% (6 out of 20) to 65\% (11 out of 17). This knowledge acquisition is also observed in Figure~\ref{fig:Q4}: the great majority of the students finalize the course relating ``supercomputers'' to systems equipped with optimized software and/or hardware. This fact is especially remarkable in the second edition, when 70\% of the students (specifically 12 out of 17) show this final opinion, in contrast to the initial 50\% (10 out of 20).

\textbf{HPC will be taken into account in the professional career path.}

If students had known more about HPC, all of them could have answered that of course it could be applied in their professional career field in the beginning, even though maybe they would need more training, but this is only initially stated by approximately half of the students (see IS answers in Figure~\ref{fig:Q6}). The other half of the students, initially thought that HPC could not be applied in their areas or they did not know how it could be done. 

\section{COURSE EVALUATION: DISCUSSION AND LESSONS LEARNT}
\label{sec:discussion}

The remaining results extracted from the surveys and the course experience itself are summarized and discussed in this section.

\textbf{Are students motivated to learn?}

Motivation is vital for learning, and university students have it. However, we (as lecturers) do not always find a way to keep it alive. In Figure~\ref{fig:Q1}, we observe that (in both editions) approximately half of the students attending the course signed up for it because they were interested in learning and felt curious. Appealing their motivation to learn and their curiosity was our intention when considering how to bring HPC to students. We believe that an in-place ``hands-on'' experience is crucial, as stated by other authors~\cite{Nersessian1989,clough2002using}.

\textbf{Choosing Raspberry Pi devices for the course and advertising their usage is attractive.}

Figure~\ref{fig:Q1} reflects that there is an interest in Raspberry Pi that motivates part of the students to attend the course. In fact, in the second edition, 30\% of the students (6 out of 20) express that knowing more about Raspberry Pi is why they decided to participate in the course. It is curious that, in the first edition, the number of students interested in Raspberry Pi was much lower, around 12\% (2 out of 17) and all of them were CS students. Despite the differences in terms of percentage between both editions, results show that finding an attractive way to present HPC is essential to initiate students in the field.

\textbf{Raspberry Pi components provide sufficient flexibility and versatility.}

Reasonable prices of Raspberry Pi and all the other components employed for the clusters offer the possibility of enabling students to group and develop their clusters independently. This lays good foundations for creativity and favors that, after completing the basic cluster configuration, they focus on what they are more interested in. For instance, hardware experiments such as interconnecting different clusters, understanding temperature consequences, or analyzing performance differences between shared and distributed memory executions.

\textbf{The approach and programming of the course are appropriate to establish fundamental knowledge about HPC.}

After the second edition of the course, we can reaffirm that a 10-hour program lets the students learn the basics of HPC, motivates them to keep/raise their HPC interest and curiosity, and encourages them to experiment with their cluster by applying it their proposals. 

\begin{figure}[htbp]
    \centering
    \includegraphics[width=0.85\linewidth]{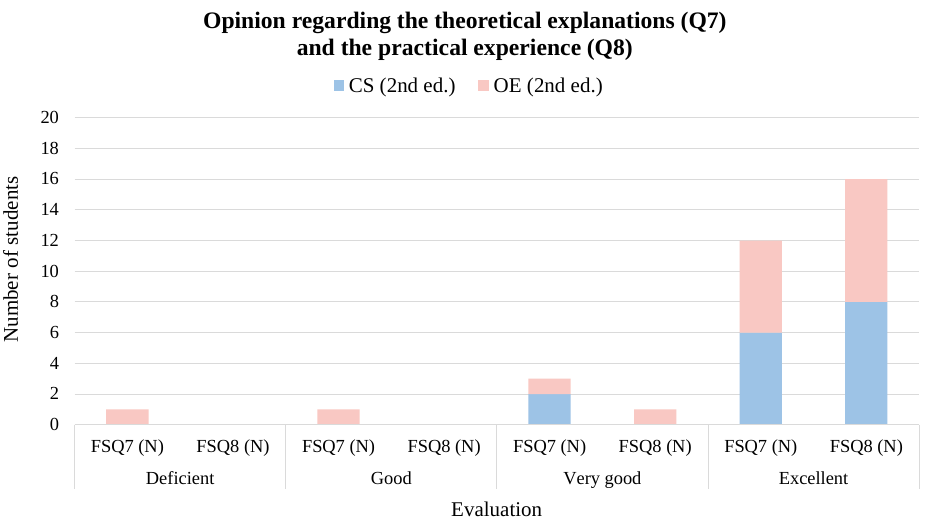}
    \caption{Answers of FSQ7 and FSQ8. Note that only second edition answers are provided, as these questions were newly included in it.}
    \label{fig:Q7-Q8}
\end{figure}

In the second edition of the course, we included two questions (Q7 and Q8) in the final survey to evaluate the degree of satisfaction with the course, whose answers are reflected in Figure~\ref{fig:Q7-Q8}. In general, we can conclude that students liked both theoretical explanations and practical activities. There are small differences between CS and OE students and we consider that it might be caused by the lack of CS knowledge that OE students have, compared to the CS ones. Besides, only one OE student considered that our empathy degree and quality of the help during practical activities were very good, instead of excellent (as all the other students stated), which makes us believe that what was not {\it very good} or {\it excellent} for certain students during the theoretical explanations was compensated afterward during the practical section of the course. Thus, we consider that the theoretical explanations should be kept as they prevent 1) the risk of losing the attention of CS students with too basic explanations, and 2) reducing the on-demand time slot because of providing longer explanations.

{\bf Students' original professional interests are kept, but now HPC is considered in those contexts.}

\begin{figure}[htbp]
    \centering
    \includegraphics[width=0.85\linewidth]{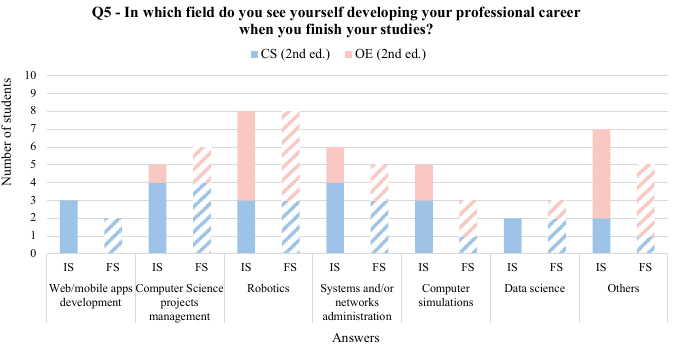}
    \caption{Answers of ISQ5 and FSQ5. Note that multiple answers where accepted for Q5.}
    \label{fig:Q5}
\end{figure}

We can observe in Figure~\ref{fig:Q5} that future professional interests are very different from one to other students, and taking the course does not modify their preferences. However, Figure~\ref{fig:Q6} reflects that, after the course, fewer students consider HPC is not applicable in their field or ignore how, and more believe that they could use it, although most express they would (naturally) need to improve their skills to be able to apply it. 

For us, it is very important to ensure and remark that acquiring basic HPC knowledge does not modify their professional interest, but enriches their perspective of how HPC can be employed to improve several applications, even though they belong to very different fields.

\textbf{Low interest in HPC among women.}

After the two editions of the course, we observe a low number of women applying to it. We consider the main reason is the fact that the number of female students in the degrees which this course was offered is reasonably low, and consequently this imbalance is also naturally present in the course. The latest data regarding the number of women enrolling in STEM degrees in the university where the course took place~\cite{Garcia.2019} show that female students represent only 11\%. In the first edition, female applicants were 7.7\% (2 out of 26), while in the second edition were 10.5\% (5 out of 48). Consequently, the number of female applicants seems to be consistent with the existing number of female students in these fields.

On the other hand, the number of final attendees does not keep consistent between both editions. While in the first edition the selected woman attended the course, in the second one none of them finally participated. We consider that we do not have enough information to make a strong statement about the reasons that lead to these results. In future editions, more information will be gathered regarding female interest.

\textbf{One-day vs. two-day course.}

One of the main differences between the first and second editions of the course was its time distribution; the first edition was split into two days of five hours each, while the second edition was carried out in one day for ten hours. This change in the format of the course was because the exam period was close by. Before the second edition of the course was published, we guessed that fewer students would apply because it was taking place on Saturday (a non-school day), and the personal effort from students would be greater, spending all day in the workshop. Surprisingly, quantitative results did not differ much from the previous edition. However, we consider them qualitatively better after analyzing the following three aspects:
\begin{itemize}
    \item The number of applicants raised significantly from the previous edition to the second, going from 26 to 48. 
    \item The number of applicants belonging to CS or OE slightly varied between the two editions. In the first edition, 58\% of the applicants were from CS, while this percentage in the second edition turned to be 50\%.
    \item The number of attendees who enrolled in the course versus the number of them that finished it was similar. In the first edition, 11\% of the attendees dropped the course before its end. In the second edition, this amount rose to 15\%.
\end{itemize}

The interest in the course increased drastically from one edition to the next one. There could be different reasons for that, but we consider that the previous knowledge about the course existence, together with offering it during a non-school day, has made an important difference.

Regarding the major of the applicants, similar numbers are obtained in both editions. In the first edition, CS students seemed to be a bit more interested than in the second one. We see this fact as another evidence that HPC is a transverse field that is widely used by many different sciences and engineering nowadays. Consequently, students from all areas may be interested in it.

The dropout rate is similar in both editions, being a bit higher in the second. We think that this may be caused by the duration of the course. Spending 10 hours on a day may be difficult for some of the attendees that may have other obligations to take care of. 

In terms of the course contents, we observed a remarkable difference that was very satisfying from our point of view. In contrast to the first edition, in the second one, some of the students that progressed faster thanks to their previous CS knowledge were able to experiment and develop more activities (see Section~\ref{subsec:MethodPart3}). This was especially observable during the on-demand phase and gives a non-negligible value that, in our humble opinion, enriches the students' experience a lot.
For this reason, we are determined to keep offering the course in this ``whole day experience'' format instead of splitting it into two sessions.

\section{FOLLOW-UP: IMPACT OF ATTENDING THE COURSE}\label{sec:follow_up}
In the first quarter of the year 2021 (respectively two and three years after the first and second edition), the authors sent a short questionnaire to the participants of both editions of the course to follow-up on the impact and influence of attending it. 
The questionnaire was composed of two questions that the students rated from $0$ (``\textit{Not at all}'') to $10$ (``\textit{Absolutely}''):
\begin{itemize}
    \item Did attending the course influence any subsequent training choices? For example, in the selection of specialization pathways for undergraduate studies, or specific training courses.
    \item Do you think that the training obtained in the course has any impact on the field in which you are currently studying or working?
\end{itemize}

\begin{figure}[ht]
    \centering
    \includegraphics[width=\textwidth]{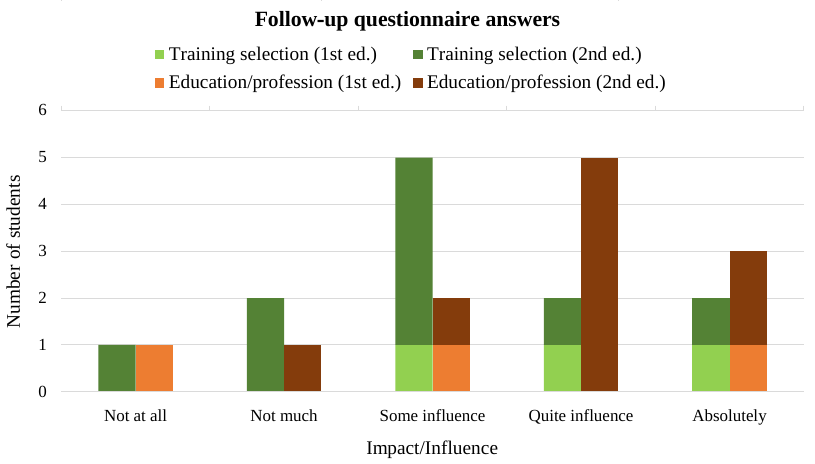}
    \caption{Answers of the follow-up questionnaire.}
    \label{fig:cuestionario}
\end{figure}

Figure~\ref{fig:cuestionario} reports the questionnaire answers, showing in orange the influence over posterior training selection by the attendees, and in green, the impact in their current studies and/or the professional environment; the light tonalities refer to the first edition and darker tones to the second one. 

In total, 12 students out of 38, answered the follow-up survey. Three of them attended the first edition, which means that only 16\% of that edition members are represented; the other nine answers belong to students from the second edition, which is a much more representative opinion, reflecting 45\% of the original attendance. Most of the students lose or stop checking their university email inbox when they conclude their studies, and that can be the reason why some of them did not reply.

In general, there is a moderate influence of the course attendance over training selection and a high impact of the acquired skills in the studies and/or professional development. The lack of HPC content in the engineering syllabus (already explained in this paper) can be the reason why the attendees did not vary much in their afterward training selection, as there are not many options related to what they learned in the course. However, as it was also stated in the introduction, nowadays engineering-related professions often require generalist HPC knowledge, and that is reflected in the high impact of the course HPC apprenticeship in the attendees' education and/or profession.

The already mentioned low representation of the first edition can justify the {\it polarized} opinions, while a most representative scope like that regarding the second edition homogenizes the data and is closer to what can be observed globally without differentiating editions. For this reason, we consider that the main conclusion that can be extracted regarding the different editions is that the attendees of the second one show a clear impact on the course in terms of training selection and mostly regarding educational/professional development.

\section{CONCLUSIONS}\label{sec:conclusions}

HPC necessity and importance are unquestionable, and we have realized that there is a lack of related content in the engineering syllabus of the UJI, particularly in CS subjects. Motivated students and an increasing necessity of HPC knowledge in the job market, moved us to propose the course. Combining the ``hands-on'' experience and the ``on-demand'' approach works in favor of creativity and motivation, and results show that the students valued positively those aspects. Regarding HPC skills and knowledge, it is seen that the number of students that define appropriately HPC and supercomputer has increased. On the other hand, the number of students that show more interest in HPC raises after the course. 

In terms of motivating the students, employing Raspberry Pi devices and giving them some freedom to experiment (in our case, through the ``on-demand'' part of the course) has been largely positive and has helped to achieve the objectives. In fact, the use of Raspberry Pi attracted the attention of the students to enroll in the course. Moreover, offering a single-day course instead of splitting it into two afternoons helps to maintain their interest.
This new format eases the understanding while linking all the new concepts we present. Besides, it facilitates to dedicate more time to the ``on-demand'' section, as no recaps are needed, and fewer questions regarding previously explained things are asked. It is also remarkable the higher number of activities that the most advanced groups can experience on the cluster, as no extra reboots or extra reconfigurations are done due to restarting everything one day after.

Regarding the students' awareness about HPC and its applicability in the real world, we can conclude that it has increased. Students know more about the impact of HPC on daily life after the course. Moreover, students seem more confident about the possibility of applying HPC in their fields, although their professional interests have not changed after the course.

Finally, we can conclude that the course had an impact not only in the short-term but also in the mid-/long-term. The follow-up questionnaire results show that, in some cases, attendees chose HPC-related subjects and courses after the proposed course. Moreover, a reasonable percentage of them are applying their HPC knowledge in their current job or field of study.

Notice that these conclusions are extracted from a small population, so they may be less accurate in a scaled-up environment. However, the course could not be carried out with more students due to the tight budget available for this activity. 



\section{FUTURE WORK}\label{sec:future-work}
We are satisfied with the results derived from completing the course and will offer subsequent editions of it.
We plan to look for funding from technological companies so we can include a small competition at the end of the course in which a general challenge is proposed, following the spirit of student cluster competitions.

Moreover, we are working on including Slurm installation and usage, as well as defining an extended collection of HPC applications arising from the different students' specific fields of interest.

Considering the current health crisis due to COVID and the limitations we are experiencing in terms of planning in-person courses, it is unavoidable for us to think of designing an extension that allows us to run everything virtually. Although the hands-on is one of the highlights of this course, the new pandemic situation is affecting and will possibly affect further editions, so we believe it is worth putting a big effort into designing ways to still allow this ``hands-on'' (though virtual) and maintaining the degree of freedom the students get in this course, which derives from the ``on-demand'' approach.

One of the authors' concerns is promoting HPC not only in university environments but also in high schools. We believe that teenagers would earlier discover CS in general and HPC in particular through experiencing a properly adapted version of this course. As a consequence, their corresponding perspective could be more adjusted to reality and less limited than nowadays. We also consider that discovering and experiencing CS at earlier ages could increase interest in this field among girls, and hopefully contribute to reducing the gender imbalance.

Lastly, we are also considering coupling all the materials that make this course possible and publishing them (extended with complementary explanations, exercises, and developments) in book format.

\section*{Acknowledgment}

The researcher Sandra Catal\'an was supported by MICINN under the project Heterogeneidad y Especializaci\'on en la Era Post-Moore, RTI2018-093684-B-I0. Roc\'io Carratal\'a-S\'aez was supported by projects CICYT TIN2014-53495-R and TIN2017-82972-R of MINECO and FEDER, project UJI-B2017-46 of UJI, and the FPU program of MECD. Sergio Iserte was supported by a postdoctoral fellowship from Valencian Region Government and European Social Fund, APOSTD/2020/026. This course was possible thanks to the funding provided by the High Performance Computing and Architectures Research Group (HPC\&A), and Computer Science Department (DICC) from UJI.
The authors also want to thank the anonymous reviewers of the work presented in~\cite{teachingOnDemand}, whose comments were very helpful to enrich the course. Likewise, authors appreciate the anonymous reviews and suggestions of the current manuscript which helped to improve the quality of the paper.

\bibliography{main}

\newpage
\appendix

\section{User guide}\label{sec:appendix}

In this appendix, we present a ``step by step'' guide detailing the tasks proposed in the course. Besides, it is also specified if the instructions must be executed on every Raspberry Pi device or only in the front-end.

\subsection{Assemble the cluster}

\begin{itemize}
\item Insert SD cards in Raspberry Pi devices (instructors need to previously prepare them to be bootable with Raspbian OS).
\item Connect the screen, the mouse and keyboard to one of the Raspberry Pi devices.
\item Connect all the Raspberry Pi devices to the switch using Ethernet cables.
\item Connect each Raspberry Pi to the energy. 
\end{itemize}

\subsection{Basic configuration of Raspbian OS (in all the devices)}

\begin{itemize}
\item Run each Raspberry Pi and perform the basic OS configurations, such as setting the right clock time, and the proper keyboard and system language. To do this, alternate the screen, the mouse and keyboard from one to the other Raspberry Pi devices.
\end{itemize}

\subsection{Further configuration of Raspbian OS (in all the devices)}

\begin{itemize}
\item Assign a hostname to each device (for example, use \texttt{nodeX} where \texttt{X} is substituted by 1, 2, 3, and 4 for each device).
\item Enable SSH. This can be done through system interfaces configuration.
\item It is highly recommended to check that geographical location is correctly set.
\item After rebooting, configure DHCP using:
\begin{itemize}
    \item interface: \texttt{eth0}
    \item static ip\_address: \texttt{192.168.0.x/24} where \texttt{x} is substituted by 1, 2, 3, and 4 for each device.
    \item static routers: \texttt{192.168.0.1}
    \item static domain\_name\_severs: \texttt{192.168.0.1}
\end{itemize}
\item Reboot the devices.
\end{itemize}

\subsection{Configure the front-end (\texttt{node1})}

\begin{itemize}
    \item In the system preferences, set hostname and enable SSH:
    \begin{itemize}
        \item Hostname: \texttt{node1}
        \item Interfaces: SSH
    \end{itemize}
    \item Reboot the system and then configure DHCP by modifying the \texttt{/etc/dhcpcd.conf} file to add:
    \begin{itemize}
        \item interface eth0
        \item static ip\_address=192.168.0.1/24
        \item static routers=192.168.0.1
        \item static domain\_name\_severs=192.168.0.1 8.8.8.8
    \end{itemize}
    \item Reboot the system or type \texttt{sudo ifconfig eth0 down \& sudo ifconfig eth0 up}.
    \item Enable WiFi.
    \item Create the file \texttt{/lib/dhcpcd/dhcpcd-hooks/60-gw} and write \texttt{route del default gw 192.168.0.1} on it.
    \item Reboot the system and overwrite the file \texttt{/etc/hosts} with:
    \begin{itemize}
        \item 192.168.0.1 	node1
        \item 192.168.0.2 	node2
        \item 192.168.0.3 	node3
        \item 192.168.0.4 	node4
    \end{itemize}
    \item Reboot the device and generate SSH keys typing \texttt{ssh-keygen}.
    \item Configure SSH in the other nodes, from \texttt{node1} by repeating three times (substituting \texttt{X} by 2, 3, and 4):
    \begin{itemize}
        \item \texttt{ssh-copy-id nodeX}
        \item \texttt{scp /etc/hosts nodeX:}
        \item \texttt{ssh nodeX sudo mv hosts /etc/hosts}
    \end{itemize}
    \item Update the system (\texttt{sudo apt-get update}) and install NFS server by typing \texttt{sudo apt-get install nfs-kernel-server}.
    \item Create a shared directory:
    \begin{itemize}
        \item \texttt{sudo mkdir /SHARED}
        \item \texttt{sudo chmod 777 /SHARED}
        \item Modify the file \texttt{/etc/exports} by adding:
        \begin{itemize}
            \item \texttt{/SHARED node2(rw, sync, no\_subtree\_check)}
            \item \texttt{/SHARED node3(rw, sync, no\_subtree\_check)}
            \item \texttt{/SHARED node4(rw, sync, no\_subtree\_check)}
        \end{itemize}
        \item \texttt{sudo exportfs -a}
    \end{itemize}
    \item Make the shared directory accessible for the other nodes (these steps can be done by entering each node using SSH, and need to be done three times):
    \begin{itemize}
        \item \texttt{sudo mkdir /SHARED}
        \item \texttt{sudo chmod 777 /SHARED}
        \item Modify the file \texttt{/etc/fstab} by adding \texttt{node1:/SHARED /SHARED nfs}
        \item \texttt{sudo mount -a}
    \end{itemize}
\end{itemize}

\subsection{Install OpenMPI}

\begin{itemize}
    \item Download OpenMPI from \url{open-mpi.org}.
    \item Decompress downloaded files.
    \item Configure OpenMPI by typping \texttt{./configure --prefix=/SHARED/OpenMPI --enable-mpirun-prefix-by-default}.
    \item Install it by typing \texttt{make \&\& make install}.
    \item Only in \texttt{node1}, modify the file \texttt{bash.rc} by adding \texttt{export PATH=/SHARED/openmpi/bin:\$PATH}.
    \item Update \texttt{bash.rc} in the remaining nodes by typping \texttt{scp .bashrc nodeX:}.
    \item It is recommended to check that OpenMPI has been successfully installed by running \texttt{mpiexec hostname}. 
\end{itemize}

\subsection{Alternatively, to check different performance rates, MPICH can also be installed following these steps:}

\begin{itemize}
    \item Download MPICH from \url{https://www.mpich.org/}.
    \item Decompress downloaded files.
    \item Configure it by typping \texttt{./configure --prefix=/SHARED/mpich --disable-f77 --disable-fc --disable-fortran}.
    \item Install it by typing \texttt{make \&\& make install}.
    \item Only in \texttt{node1}, modify the file \texttt{bash.rc} by adding \texttt{export PATH=/SHARED/mpich:\$PATH}.
    \item Update \texttt{bash.rc} in the remaining nodes by typping \texttt{scp .bashrc nodeX:}.
\end{itemize}

\subsection{Install LINPACK}
In order to install, configure and  execute LINPACK, students are encourage to  follow the LINPACK installation guide, and so they get familiar with following an installation guide by themselves and facing system administrators issues.
However, instructors will help them with the use of configuration wizards such as \url{https://www.advancedclustering.com/act_kb/tune-hpl-dat-file} or \url{http://hpl-calculator.sourceforge.net}.

\subsection{Install LAMMPS}
In the case of LAMMPS installation, students are led to use the official user guide. Furthermore, apart from installing the basic version of LAMMPS, students are also motivated to go further and install the OpenMP extensions from \url{https://lammps.sandia.gov/doc/Build_extras.html#user-omp}. 

With LAMMPS building versions (MPI and OpenMP) students will be asked to carry out the scalability and performance evaluation using different approaches such as threads, processes and their combination.

\end{document}